\documentclass[]{article}

\let\realgeq\geq

\usepackage{graphicx}
\usepackage[latin1]{inputenc}
\usepackage{amssymb,amsfonts}
\usepackage{theorem}

\newtheorem{theorem}{Theorem}

\newtheorem{corollary}[theorem]{Corollary}

{\theorembodyfont{\rmfamily}
\newtheorem{definition}[theorem]{Definition}}
{\theorembodyfont{\rmfamily}
\newtheorem{example}[theorem]{Example}}

\newtheorem{proposition}[theorem]{Proposition}
{\theorembodyfont{\rmfamily}
}

\newenvironment{proof}[1][Proof]{\noindent\textbf{#1.} }{\ \rule{0.5em}{0.5em}}

\begin{document}

\title{Stability between foliations in general relativity}

\author{V. J. Bol\'os \\{\small Dpto. Matem\'aticas, Facultad de Ciencias, Universidad de
Extremadura.}\\ {\small Avda. de Elvas s/n. 06071--Badajoz,
Spain.}\\ {\small e-mail\textup{: \texttt{vjbolos@unex.es}}}}

\maketitle

\begin{abstract}
The aim of this paper is to study foliations that remain
invariable by parallel transports along the integral curves of
vector fields of another foliations. According to this idea, we
define a new concept of stability between foliations. A particular
case of stability (called regular stability) is studied, giving a
useful characterization in terms of the Riemann curvature tensor.
This characterization allows us to prove that there are no
regularly self-stable foliations of dimension greater than $1$ in
Schwarzschild and Robertson-Walker space-times. Finally, we study
the existence of regularly self-stable foliations in other
space-times, like $pp$-wave space-times.
\end{abstract}

\section{Introduction}

During the last decades, applications of foliations to theoretical
physics have been considerably increased \cite{Do}. At the
sixties, J. M. Souriau introduced foliations associated to
elementary particles to study their evolutions in the
Min\-kows\-ki space-time \cite{Soa}. Later, the use of foliated
manifolds has provided very good results in relativity and quantum
mechanics \cite{Mo}. For instance, the symplectic bundle structure
allows us to enclose a space-time, a dynamical system and its
evolution space in the same mathematical structure. In this way, a
foliation describes the evolution of the dynamical system
\cite{Gu,Lib}. These facts have motivated us to study some general
properties of foliations.

In this paper, we analyze distributions and foliations that remain
invariable by parallel transports. If a foliation is conserved by
parallel transports along the integral curves of its vector
fields, then this foliation satisfies a {\it motion law}; in this
case it can be proved that its leaves are totally geodesic
\cite{Lia,Lic}. However, we can obtain more general properties
using parallel transports along the integral curves of vector
fields of another foliation. For example, if a foliation is
conserved by parallel transports along world lines of a congruence
of observers, then they observe the leaves of the foliation as
invariable along their evolution, and it is interesting to study.
According to this idea, we define a new concept of {\it stability}
between foliations in Section \ref{sec1}. A particular case of
stability (called {\it regular stability}) is studied in Section
\ref{sec2}, giving a useful characterization in Theorem \ref{T1}.
This result allows us to prove that there are no {\it regularly
self-stable} foliations of dimension greater than $1$ in
Schwarzschild and Robertson-Walker space-times, but there exist
foliations of this kind in other space-times. Finally, in Section
\ref{sec3}, we study the existence of regularly self-stable
foliations in $pp$-wave space-times.

\section{Stability}
\label{sec1}

We work on a $n$-dimensional space-time manifold $\mathcal{M}$
(although all results and proofs can be generalized to any
manifold with a torsion-free metric connection) and we denote the
Levi-Civita connection by $\nabla $. We use the convention that
$\hbox{\rm span}\left( X_1,\ldots ,X_p\right) $ denotes the
subbundle spanned by the vector fields $X_1,\ldots ,X_p$, and it
is called \textit{distribution}. Usually, a distribution of
dimension $p$ is called a $p$-distribution. All bases of
distributions are local. A distribution that has an integral
submanifold (leaf) in every point is a {\it foliation}. We say
that a foliation is a \textit{flat foliation} if its leaves are
flat submanifolds, and we say that a foliation is a
\textit{totally geodesic foliation} if its leaves are totally
geodesic submanifolds.

In previous works \cite{Lib,Lia,Lic}, the concept of {\it motion
law} was introduced using foliations: let $\Omega $ be a
foliation, $X$ a vector field of $\Omega $, $c$ a maximal integral
curve of $X$ and
\[
\tau _{t}^{c}:T_{c\left( 0\right) }\mathcal{M}\longrightarrow
T_{c\left( t\right) }\mathcal{M}
\]
the parallel transport along $c\left( t\right) $, for all $t\in
I$, where $I$ is the domain of $c$. Then, $\Omega $ verifies a
{\it motion law} if
\[
\tau _{t}^{c}\Omega \left( c\left( 0\right) \right) =\Omega \left(
c\left( t\right) \right) ,\quad t\in I.
\]
This motion law is equivalent to say that $\Omega $ is a totally
geodesic foliation. Intuitively, the curvature of the leaves has
to ``adapt'' to the curvature of the space-time. In Definition
\ref{def1} we show how to generalize this intuitive idea.

\begin{definition}
\label{def1} Let $\Omega ,\Omega '$\ be two distributions. We will
say that $\Omega $ is {\it stable with respect to} $\Omega '$, and
we will denote it by $\nabla _{\Omega '}\Omega \subset \Omega $,
if
\[
\nabla _{Y}X\in \Omega
\]
for all vector fields $X\in \Omega $, $Y\in \Omega '$.

Particularly, if $\Omega =\Omega '$ we will say that $\Omega $ is
{\it self-stable}.
\end{definition}

Clearly, a distribution $\Omega $ is self-stable if and only if it
is a totally geodesic foliation. Note that if $\Omega $ is a
self-stable distribution, then $\left[ X,Y\right] =\nabla
_{X}Y-\nabla _{Y}X\in \Omega $ for all $X,Y\in \Omega $. So,
$\Omega $ is involutive and hence, by Frobenius' Theorem, it is
integrable. In consequence, a self-stable distribution is in fact
a totally geodesic foliation.

In order to know if $\Omega $ is stable with respect to $\Omega '$
it is sufficient to check that, given $\left\{ X_{i}\right\}
_{i=1}^{p},\left\{ Y_{j}\right\} _{j=1}^{q}$ some arbitrary bases
of $\Omega $\ and $\Omega '$\ respectively, the following
conditions hold:
\begin{equation}
\nabla _{Y_{j}}X_{i}\in \Omega ,\qquad \left\{
\begin{array}{c}
i=1,\ldots ,p, \\
j=1,\ldots ,q.
\end{array}
\right.  \label{cbases}
\end{equation}
Besides, conditions (\ref{cbases}) show that any span of vector
fields of $\Omega $ is conserved by parallel transports along the
integral curves of vector fields of $\Omega '$.

Sometimes it is easier to deal with the orthogonal distribution of
$\Omega $ (denoted $\Omega ^{\bot }$) instead of dealing with
$\Omega $. In these cases, Proposition \ref{p100} is very useful.

\begin{proposition}
\label{p100} Let $\Omega ,\Omega ^{\prime}$ be two distributions.
Then $\Omega $ is stable with respect to $\Omega ^{\prime}$ if and
only if $\Omega ^{\bot }$ is stable with respect to $\Omega
^{\prime}$; {\it i.e.}
\[
\nabla _{\Omega ^{\prime}}\Omega \subset \Omega
\Longleftrightarrow \nabla _{\Omega ^{\prime}}\Omega ^{\bot
}\subset \Omega ^{\bot }.
\]
\end{proposition}

\begin{proof}
It is known \cite{He} that for all triplet of vector fields
$X,Y,Z$ in a pseudo-Riemannian manifold $\mathcal{M}$ with metric
$g$ and connection $\nabla $, we have
\begin{equation}
Zg\left( X,Y\right) =g\left( \nabla _{Z}X,Y\right) +g\left(
X,\nabla _{Z}Y\right) .  \label{l2.22}
\end{equation}

Necessary condition: let $\Omega ,\Omega '$ be two distributions
such that $\nabla _{\Omega '}\Omega \subset \Omega $. Given three
arbitrary vector fields $X\in \Omega $, $Y\in \Omega ^{\bot }$,
$Z\in \Omega '$, by (\ref{l2.22}) we have $0=g\left( X,\nabla
_{Z}Y\right) $. So $\nabla _{Z}Y\in \Omega ^{\bot }$, and then
$\nabla _{\Omega '}\Omega ^{\bot }\subset \Omega ^{\bot }$.

The proof of the sufficient condition is analogous.
\end{proof}

Proposition \ref{p100} says that $\Omega $ and $\Omega ^{\bot }$
have the same behaviour in relation to stability. So, given a
distribution $\Omega $, we can study the stability of $\Omega $
through the stability of $\Omega ^{\bot }$. This is very useful
when $\Omega $ is a $\left( n-1\right) $-distribution, since
$\Omega ^{\bot }$ is a $1$-distribution and the study of the
stability becomes easier. Moreover, if $\Omega $ is a lightlike
$\left( n-1\right) $-distribution, then $\Omega ^{\bot }$ is the
span of a lightlike vector field of $\Omega $. In this particular
case, the leaves of $\Omega $ are interpreted as wave fronts and
the integral curves of $\Omega ^{\bot }$ represent the world lines
of a congruence of massless particles. Hence, Proposition
\ref{p100} can be regarded as a ``wave-particle duality'' result.

\section{Regular stability}
\label{sec2}

We are going to introduce a special kind of stability, called {\it
regular stability}.

\begin{definition}
Let $\Omega ,\Omega '$\ be two distributions. We will say that
$\Omega $ is {\it regularly stable with respect to} $\Omega '$,
and we will denote it by $\nabla _{\Omega '}\Omega =0$, if there
exists a basis $\left\{ X_{i}\right\} _{i=1}^{p}$\ of $\Omega $
such that
\[
\nabla _{Y}X_{i}=0\qquad i=1,\ldots ,p,
\]
for all vector field $Y\in \Omega '$. In this case we will say
that $\left\{ X_{i}\right\} _{i=1}^{p}$\ is a {\it regularly
stable basis of} $ \Omega $ {\it with respect to} $\Omega '$.

Particularly, if $\Omega =\Omega ^{\prime}$ we will say that
$\Omega $ is {\it regularly self-stable} and $\left\{
X_{i}\right\} _{i=1}^{p}$ is a {\it regularly self-stable basis
of} $\Omega $.
\end{definition}

Given a regularly stable basis of $\Omega $ with respect to
$\Omega '$, its vector fields are conserved by parallel transports
along the integral curves of vector fields of $\Omega '$. But only
some bases of $\Omega $ have this property.

It is clear that any subset of vector fields of a regularly
self-stable basis spans a regularly self-stable foliation and,
obviously, it is a regularly self-stable basis of this foliation.
Particularly, for dimension 1, we obtain that the vector fields of
a regularly self-stable basis are geodesic.

To illustrate the concept of regular stability, let us see the
following example.

\begin{example}
In spherical coordinates $\left( t,r,\theta ,\varphi \right) $ the
metrics of Schwarzschild and Robert\-son-Walker are given by
\[
ds^{2}=\frac{1}{a_{\mathrm{S}}}dr^{2}+r^{2}\left( d\theta
^{2}+\sin ^{2}\theta d\varphi ^{2}\right) -a_{\mathrm{S}}dt^{2},
\]
\[
ds^{2}=\frac{F^{2}}{a_{\mathrm{RW}}^{2}}\left( dr^{2}+r^{2}\left(
d\theta ^{2}+\sin ^{2}\theta d\varphi ^{2}\right) \right) -dt^{2},
\]
respectively, where $a_{\mathrm{S}}:=1-\frac{2m}{r}$, $F=F\left(
t\right) \realgeq 0$, $a_{\mathrm{RW}} :=\left(
1+\frac{1}{4}kr^{2}\right) $ and $k=-1,0,1$.

Let us consider the $2$-foliations
\[
\Omega :=\hbox{\rm span}\left( \frac{\partial }{\partial \theta
},\frac{\partial }{\partial \varphi }\right) ,\quad \Omega
':=\hbox{\rm span}\left( \frac{\partial }{\partial
r},\frac{\partial }{\partial t}\right) .
\]
The leaves of $\Omega $ are surfaces with $r$ and $t$ constant
({\it i.e.} spatial 2-spheres centered on the origin), and the
leaves of $\Omega '$ are surfaces with $\theta $ and $\varphi $
constant. It is easy to prove that $\nabla _{\Omega '}\Omega =0$
in both space-times. The following bases of $\Omega $
\[
\left\{ \frac{1}{r}\frac{\partial }{\partial \theta },\quad
\frac{1}{r}\frac{\partial }{\partial \varphi }\right\} ,\quad
\left\{ \frac{a_{\mathrm{RW}}}{Fr}\frac{\partial }{\partial \theta
},\quad \frac{a_{\mathrm{RW}}}{Fr}\frac{\partial }{\partial
\varphi }\right\} ,
\]
are regularly stable with respect to $\Omega '$ in Schwarzschild
and Robertson-Walker space-times respectively.
\end{example}

Next, we are going to study the relationships between two
regularly stable bases of the same distribution $\Omega $.

\begin{proposition}
\label{stat} Let $\Omega ,\Omega '$ be two distributions such that
$\nabla _{\Omega '}\Omega =0$, and let $\left\{ X_{i}\right\}
_{i=1}^{p}$  be a regularly stable basis of $\Omega $ with respect
to $\Omega '$. Then, $\left\{ \overline{X}_{i}\right\} _{i=1}^{p}$
is another regularly stable basis of $\Omega $ with respect to
$\Omega '$ if and only if there exists a family of functions
$\left\{ \alpha _{i}^{j}\right\} _{i,j=1}^{p}$ such that
\begin{itemize}
\item $\det \alpha _{i}^{j}\neq 0$, \item $\overline{X}_{i}=\alpha
_i^jX_j$ for all $i=1,\ldots ,p$, \item $Y\left( \alpha
_{i}^{j}\right) =0$, for all $i,j=1,...,p$, and for all $Y\in
\Omega '$ ({\it i.e.} $\left\{ \alpha _{i}^{j}\right\}
_{i,j=1}^{p}$ is a family of constant functions for $\Omega '$).
\end{itemize}
\end{proposition}

\begin{proof}
Necessary condition: let us suppose that $\left\{
\overline{X}_{i}\right\} _{i=1}^{p}$ is a regularly stable basis
of $\Omega $ with respect to $\Omega '$. Then, it is clear that
there exists a family of functions $\left\{ \alpha
_{i}^{j}\right\} _{i,j=1}^{p}$ such that $\det \alpha _{i}^{j}\neq
0$, and $\overline{X}_{i}=\alpha _i^jX_j$ for all $i=1,\ldots ,p$.
Let $Y$ be an arbitray vector field in $\Omega '$. Then
\begin{equation}
0=\nabla _{Y}\overline{X}_{i}=\nabla _{Y}\left( \alpha
_{i}^{j}X_{j}\right) =Y\left( \alpha _{i}^{j}\right) X_{j}+\alpha
_{i}^{j}\nabla _{Y}X_{j},\qquad i=1,...,p.  \label{fp1}
\end{equation}
Since $\nabla _{Y}X_{j}=0$ for all $j=1,...,p$, by (\ref{fp1}) we
have $Y\left( \alpha _{i}^{j}\right) X_{j}=0$ for all $i=1,...,p$
and then $Y\left( \alpha _{i}^{j}\right) =0$ for all
$i,j=1,...,p$.

Sufficient condition: it is clear that $\left\{
\overline{X}_{i}\right\} _{i=1}^{p}$ is another basis of $\Omega
$. Moreover, given $Y\in \Omega '$ we have
\begin{equation}
\nabla _{Y}\overline{X}_{i}=\nabla _{Y}\left( \alpha
_{i}^{j}X_{j}\right) =Y\left( \alpha _{i}^{j}\right) X_{j}+\alpha
_{i}^{j}\nabla _{Y}X_{j},\qquad i=1,...,p.  \label{fp2}
\end{equation}
Since $\nabla _{Y}X_{j}=0$ for all $j=1,...,p$, and $Y\left(
\alpha _{i}^{j}\right) =0$ for all $i,j=1,...,p$, by (\ref{fp2})
we have $\nabla _{Y}\overline{X}_{i}=0$ for all $i=1,...,p$,
concluding the proof.
\end{proof}

Proposition \ref{stat} assures us uniqueness, up to constant
functions for $\Omega '$, of regularly stable bases of $\Omega $
with respect to $\Omega '$. Moreover, using this result, given a
regularly stable basis of $\Omega $ with respect to $\Omega '$, we
can construct all the regularly stable bases of $\Omega $ with
respect to $\Omega '$.

The main result of this paper is given in the next theorem,
showing an operational condition for the equivalence between
stability and regular stability in terms of the Riemann curvature
tensor $R$. This condition is very useful because the study of
regular stability is easier than the study of stability in
general.

\begin{theorem}
\label{T1} Let $\Omega $ and $\Omega '$ be a $p$-distribution and
a $q$-foliation respectively such that $\nabla _{\Omega '}\Omega
\subset \Omega $. Then, $\nabla _{\Omega '}\Omega =0$ if and only
if $R\left( Y,Z\right) X=0$ for all $X\in \Omega $ and for all
$Y,Z\in \Omega '$.
\end{theorem}

\begin{proof}
Let $\left\{ X_{i}\right\} _{i=1}^{p},\left\{ Y_{j}\right\}
_{j=1}^{q}$ be two bases of $\Omega $ and $\Omega '$ respectively,
where $p=\dim \Omega $ and $q=\dim \Omega '$. Then, there exist
some functions $h_{jk}^{i},$ where $i,k=1,\ldots ,p$ and
$j=1,\ldots ,q$ such that
\begin{equation}
\nabla _{Y_{j}}X_{k}=h_{jk}^{i}X_{i},\qquad \left\{
\begin{array}{c}
k=1,\ldots ,p, \\
j=1,\ldots ,q.
\end{array}
\right.  \label{1}
\end{equation}
Since $\Omega '$ is a foliation, we can suppose that $Y_{j}=\frac{
\partial }{\partial x^{j}}$ for $j=1,\ldots q$, where $\left( x^{1},\ldots
,x^{n}\right) $ is a flat chart for $\Omega '$.

Let us state the eqs. $\nabla _{j}\left( y^{i}X_{i}\right) =0$ for
$j=1,\ldots ,q$, where $\nabla _{j}$ denotes $\nabla
_{\frac{\partial }{\partial x^{j}}}$ and $y^{i}$ are unknown
functions for $i=1,\ldots ,p$. By using (\ref{1}), we have
\begin{equation}
\left( \frac{\partial y^{i}}{\partial
x^{j}}+y^{k}h_{jk}^{i}\right) X_{i}=0,\qquad j=1,\ldots ,q.
\label{2}
\end{equation}
Since $\left\{ X_{i}\right\} _{i=1}^{p}$ is a linearly independent
family of vector fields, expression (\ref{2}) becomes
\begin{equation}
\frac{\partial y^{i}}{\partial x^{j}}+y^{k}h_{jk}^{i}=0,\qquad
\left\{
\begin{array}{c}
i=1,\ldots ,p, \\
j=1,\ldots ,q.
\end{array}
\right.  \label{2b}
\end{equation}

The system (\ref{2b}) is formed by $q$ first order homogeneous
linear sub-systems with $p$ differential equations and $p$ unknown
functions each one. In each sub-system, it appears only one
differential operator $\frac{\partial }{\partial x^{j}}$ for
$j=1,\ldots ,q$. By a Frobenius' Theorem (see \cite{Hi}), we have
that (\ref{2b}) has non-zero solutions if and only
if some compatibility conditions (between the $q$ sub-systems that form (\ref%
{2b})) are satisfied. These conditions are known as the
\textquotedblleft cross-derivatives conditions\textquotedblright\
and are built imposing that the cross-derivatives of the functions
$y^{i}$ are equal:
\[
\left.
\begin{array}{c}
\frac{\partial }{\partial x^{l}}\left( \frac{\partial
y^{i}}{\partial x^{j}}\right) =\frac{\partial }{\partial
x^{l}}\left( -y^{k}h_{jk}^{i}\right) =-\frac{\partial
y^{k}}{\partial x^{l}}h_{jk}^{i}-y^{k}\frac{\partial
h_{jk}^{i}}{\partial x^{l}} \\
\frac{\partial }{\partial x^{j}}\left( \frac{\partial
y^{i}}{\partial x^{l}}\right) =\frac{\partial }{\partial
x^{j}}\left( -y^{k}h_{lk}^{i}\right) =-\frac{\partial
y^{k}}{\partial x^{j}}h_{lk}^{i}-y^{k}\frac{\partial
h_{lk}^{i}}{\partial x^{j}}
\end{array}
\right\} \Longrightarrow
\]
\begin{equation}
\Longrightarrow \frac{\partial y^{k}}{\partial
x^{l}}h_{jk}^{i}+y^{k}\frac{\partial h_{jk}^{i}}{\partial
x^{l}}=\frac{\partial y^{k}}{\partial
x^{j}}h_{lk}^{i}+y^{k}\frac{\partial h_{lk}^{i}}{\partial
x^{j}},\qquad \left\{
\begin{array}{c}
i=1,\ldots ,p, \\
j,l=1,\ldots ,q.
\end{array}
\right.  \label{condtemp}
\end{equation}
Taking into account (\ref{2b}) and changing indexes, from
(\ref{condtemp}) we obtain
\[
\left( h_{lk}^{m}h_{jm}^{i}-h_{jk}^{m}h_{lm}^{i}+\frac{\partial
h_{lk}^{i}}{\partial x^{j}}-\frac{\partial h_{jk}^{i}}{\partial
x^{l}}\right) y^{k}=0,\qquad \left\{
\begin{array}{c}
i=1,\ldots ,p, \\
j,l=1,\ldots ,q.
\end{array}
\right.
\]
So, a necessary and sufficient condition for the system (\ref{2b})
to have non-zero solutions is
\begin{equation}
h_{lk}^{m}h_{jm}^{i}-h_{jk}^{m}h_{lm}^{i}+\frac{\partial
h_{lk}^{i}}{\partial x^{j}}-\frac{\partial h_{jk}^{i}}{\partial
x^{l}}=0,\qquad \left\{
\begin{array}{c}
i,k=1,\ldots ,p, \\
j,l=1,\ldots ,q.
\end{array}
\right.  \label{cond}
\end{equation}
Moreover, if (\ref{cond}) is satisfied, the set of solutions of
(\ref{2b}) form a vector space of dimension $p$ (see \cite{Hi}),
{\it i.e.} there exists a family of differentiable functions
$\left\{ f_{i}^{k}\right\} _{i,k=1}^{p}$ such that $\det
f_{i}^{k}\neq 0$ and any solution of (\ref{2b}) has the form
\[
y^{k}=C^{i}f_{i}^{k},\qquad k=1,\ldots ,p,
\]
where $\left\{ C^{i}\right\} _{i=1}^{p}$ are parameter functions
({\it i.e.} $\frac{\partial C^{i}}{\partial x^{j}}=0$ for
$i=1,\ldots ,p$ and $j=1,\ldots ,q$). Hence, $\left\{
f_{i}^{k}X_{k}\right\} _{i=1}^{p}$ is a regularly stable basis of
$\Omega $ with respect to $\Omega '$, and so $\nabla _{\Omega
'}\Omega =0$ if and only if (\ref{cond}) is satisfied.

So, we have to prove that (\ref{cond}) is equivalent to $R\left(
Y,Z\right) X=0$ for all $X\in \Omega $ and for all $Y,Z\in \Omega
'$. In fact, from the linearity of the Riemann curvature tensor,
we have to prove that (\ref{cond}) is equivalent to $R\left(
\frac{\partial }{\partial x^{j}},\frac{\partial }{\partial
x^{l}}\right) X_{i}=0$ for all $i=1,\ldots ,p$ and $j,l=1,\ldots
q$:
\[
R\left( \frac{\partial }{\partial x^{j}},\frac{\partial }{\partial
x^{l}}\right) X_{i}=0\Longleftrightarrow \nabla _{j}\nabla
_{l}X_{i}-\nabla _{l}\nabla _{j}X_{i}=0,\qquad \left\{
\begin{array}{c}
i=1,\ldots ,p, \\
j,l=1,\ldots ,q.
\end{array}
\right.
\]
Applying (\ref{1}) we have
\[
\Longleftrightarrow \nabla _{j}\left( h_{li}^{k}X_{k}\right)
-\nabla _{l}\left( h_{ji}^{k}X_{k}\right) =0
\]
\[
\Longleftrightarrow h_{li}^{k}\nabla _{j}X_{k}+\frac{\partial
h_{li}^{k}}{\partial x^{j}}X_{k}-h_{ji}^{k}\nabla
_{l}X_{k}-\frac{\partial h_{ji}^{k}}{\partial x^{l}}X_{k}=0
\]
\[
\Longleftrightarrow h_{li}^{k}h_{jk}^{m}X_{m}+\frac{\partial
h_{li}^{k}}{\partial
x^{j}}X_{k}-h_{ji}^{k}h_{lk}^{m}X_{m}-\frac{\partial h_{ji}^{k}}{
\partial x^{l}}X_{k}=0,\qquad \left\{
\begin{array}{c}
i=1,\ldots ,p, \\
j,l=1,\ldots ,q.
\end{array}
\right.
\]
Changing indexes,
\[
\Longleftrightarrow \left(
h_{lk}^{m}h_{jm}^{i}-h_{jk}^{m}h_{lm}^{i}+\frac{\partial
h_{lk}^{i}}{\partial x^{j}}-\frac{\partial h_{jk}^{i}}{\partial
x^{l}}\right) X_{i}=0
\]
\begin{equation}
\Longleftrightarrow
h_{lk}^{m}h_{jm}^{i}-h_{jk}^{m}h_{lm}^{i}+\frac{\partial
h_{lk}^{i}}{\partial x^{j}}-\frac{\partial h_{jk}^{i}}{\partial
x^{l}}=0,\qquad \left\{
\begin{array}{c}
i,k=1,\ldots ,p, \\
j,l=1,\ldots ,q.
\end{array}
\right.  \label{cond2}
\end{equation}
Since expressions (\ref{cond}) and (\ref{cond2}) are the same, we
conclude the proof.
\end{proof}

Next, we are going to give some useful corollaries of Theorem
\ref{T1} related to some interesting cases.

\begin{corollary}
\label{col1} Let $\Omega $ and $\Omega '$ be a $p$-distribution
and a $q$-foliation respectively such that $\nabla _{\Omega
'}\Omega \subset \Omega $.

\begin{description}
\item[(i)] In a flat space-time (Minkowski) we have that $\nabla
_{\Omega '}\Omega \subset \Omega $ if and only if $\nabla _{\Omega
'}\Omega =0$.

\item[(ii)] If $q=1$ we have that $\nabla _{\Omega '}\Omega
\subset \Omega $ if and only if $\nabla _{\Omega '}\Omega =0$.
\end{description}
\end{corollary}

In these cases, the study of stability becomes the study of
regular stability. This fact simplifies remarkably the problem.

\begin{figure}[tbp]
\begin{center}
\includegraphics[width=0.4\textwidth]{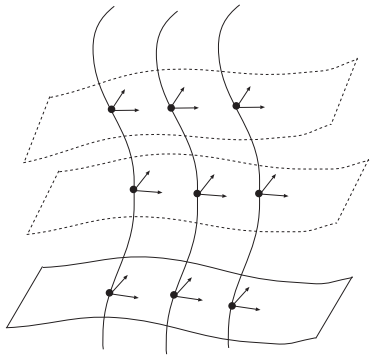}
\end{center}
\caption{If $\Omega $ is a $(n-1)$-foliation and $\Omega
'=\mathrm{span}(Y)$ is a 1-foliation such that $\nabla _{\Omega
'}\Omega \subset \Omega $, then $\nabla _{\Omega '}\Omega =0$ by
Corollary \ref{col1} (ii). Moreover, if $Y$ is not in $\Omega $,
then it is possible to reconstruct the entire foliation $\Omega $
from only one leaf, by means of parallel transports of a regularly
stable basis of $\Omega $ with respect to $\Omega '$ along the
integral curves of $Y$.} \label{fig1}
\end{figure}

Let us suppose that $\Omega '=\mathrm{span}(Y)$. According to
Corollary \ref{col1} (ii), there exist regularly stable bases of
$\Omega $ with respect to $\Omega '$, {\it i.e.} bases of $\Omega
$ whose vector fields are conserved by parallel transports along
the integral curves of $Y$. If $\Omega $ is a $(n-1)$-foliation
and $Y$ is a vector field which is not contained in $\Omega $,
then it is possible to reconstruct the entire foliation from only
one leaf of $\Omega $, by means of parallel transports of a
regularly stable basis of $\Omega $ with respect to $\Omega '$
along the integral curves of $Y$ (see fig. \ref{fig1}). Moreover,
if $Y$ is a future-pointing timelike vector field, then its
integral curves represent observers and therefore, each observer
detects $\Omega $ as invariable along its evolution ({\it i.e.}
along its world line), as we show in the next example.

\begin{example}
In the Schwarzschild space-time with spherical coordinates,
$U:=\frac{\partial }{\partial t}$ is a future-pointing timelike
vector field, whose integral curves represent stationary
observers. We are going to find all the lightlike 3-foliations
that are stable with respect to $\hbox{\rm span}\left( U\right) $,
{\it i.e.} all the light waves that are observed as invariable by
any stationary observer. If we don't take into account Corollary
\ref{col1} (ii), this becomes a hard work. But, applying this
result, we only have to find the lightlike 3-foliations that are
regularly stable with respect to $\hbox{\rm span}\left( U\right)
$. We obtain only two lightlike 3-foliations:
\[
\hbox{\rm span}\left( \pm \frac{\partial }{\partial
t}+\frac{\partial }{\partial r}a_{\mathrm{S}},\frac{\partial
}{\partial \theta },\frac{\partial }{\partial \varphi }\right) .
\]
The leaves of these foliations are spheres expanding and
contracting respectively at the speed of light. So, all the
observers represented by the integral curves of $U$ detect these
foliations as invariable along their evolutions. But the most
remarkable fact is that there are no other foliations with this
property.
\end{example}

\begin{corollary}
\label{col2} Let $\Omega $ be a self-stable $p$-foliation. Then
$\Omega $ is regularly self-stable if and only if
\begin{equation}
\label{corii} R\left( Y,Z\right) X=0,
\end{equation}
for all $X,Y,Z\in \Omega $.
\end{corollary}

It is important to remark that condition (\ref{corii}) does not
imply that $\Omega $ is a flat foliation. For example, in the
Minkowski space-time, any foliation satisfies (\ref{corii}) but it
is not necessarily flat. However, in general, if $\Omega $ is
totally geodesic ({\it i.e.} it is self-stable) then
$\overline{R}\equiv R$ and vice versa, in the sense that
$\overline{R}\left( \overline{Y},\overline{Z}\right)
\overline{X}=R\left( Y,Z\right) X$ where $\overline{R}$ is the
Riemann curvature tensor of the metric $\overline{g}$ induced in
the leaves of $\Omega $, $i:\mathcal{M}\left( \Omega \right)
\longrightarrow \mathcal{M}$ is the canonical inclusion and
$X=i_{\ast }\overline{X},Y=i_{\ast }\overline{Y},Z=i_{\ast
}\overline{Z}$. So, if $\Omega $ is self-stable, then it is flat
if and only if (\ref{corii}) is satisfied.

A foliation is regularly self-stable if and only if it is totally
geodesic and flat, and hence, the regular self-stability
generalizes the concept of flat wave fronts, introduced by J.M.
Souriau in the Minkowski space-time \cite{Sob}. We will discuss
this fact deeply in Section \ref{sec5}.

\begin{example}
In the Schwarzschild and Robertson-Walker space-times, we can
prove easily that there are not any distribution $\Omega $ of
dimension greater than 1 such that $R\left( Y,Z\right) X=0$ for
all $X,Y,Z\in \Omega $. So, by Corollary \ref{col2}, there are not
any regularly self-stable foliations of dimension greater than 1.
But, of course, there exist self-stable foliations, for example,
in spherical coordinates, the timelike $2$-foliation
\[
\hbox{\rm span}\left( \frac{\partial }{\partial r},\frac{\partial
}{\partial t}\right)
\]
is a self-stable foliation whose leaves are surfaces with $\theta
$ and $\varphi $ constant.
\end{example}

We will show, in Section \ref{sec3}, that there exist regularly
self-stable foliations of dimension greater than 1 in $pp$-wave
space-times. Moreover, we can find these kinds of foliations in
other space-times, as we show in the next example.

\begin{example}
If we consider the metric $ds^2=-\frac{1}{z}dt^2+dx^2+dy^2+zdz^2$
in the open set $\left\{ \left( t,x,y,z\right) : z>0\right\} $,
then the Einstein tensor is positive definite. So it is a valid
non-flat space-time. The spacelike 2-foliation given by
\[
\hbox{\rm span}\left( \frac{\partial }{\partial x},\frac{\partial
}{\partial y}\right)
\]
is self-stable and satisfies (\ref{corii}). So, by Corollary
\ref{col2}, we obtain that this foliation is a regularly
self-stable foliation. A regularly self-stable basis is given by
$\left\{ \frac{\partial }{\partial x},\frac{\partial }{\partial
y}\right\} $.
\end{example}

\section{Examples of regularly self-stable foliations in $pp$-wave
space-times} \label{sec3}

It is known \cite{Ma} that, in standard coordinates $\left(
u,v,y,z\right) $, a $pp$-wave metric can be expressed by
$ds^{2}=dy^{2}+dz^{2}-2Hdu^{2}-2dudv$, where $u,v$ are the
retarded and the advanced time coordinates respectively, and
$H=H\left( u,y,z\right)$. According to \cite{Ma}, in a $pp$-wave
space-time, the lightlike hypersurfaces with $u$ constant are
leaves of a lightlike $3$-foliation $\Omega $ given by
\[
\Omega :=\hbox{\rm span}\left( \frac{\partial }{\partial
v},\frac{\partial }{\partial y},\frac{\partial }{\partial
z}\right) ,
\]
and its leaves are called {\it plane-fronted gravitational waves
with parallel rays}. The foliation $\Omega $ is self-stable and
flat, {\it i.e.} $R \left( Y,Z \right) X=0$ for all $X,Y,Z\in
\Omega $. By applying Corollary 2, we obtain that $\Omega $ is a
regularly self-stable foliation. Then there exists a basis
$\left\{ X_{i}\right\} _{i=1}^{3}$\ of $\Omega $\ such that
$\nabla _{X}X_{i}=0$, $i=1,2,3$, for all vector field $X\in \Omega
$.

This fact gives us a new geometrical perspective of the
plane-fronted gravitational waves with parallel rays, because it
ensures us explicitly the existence of this kind of bases of
$\Omega $ that remain invariable under parallel transports along
the integral curves of vector fields of $\Omega $.

For example, a regularly self-stable basis of $\Omega $ is given
by $\left\{ \frac{\partial }{\partial v},\frac{\partial }{\partial
y}, \frac{\partial }{\partial z}\right\} $, and we can use
Proposition \ref{p100} to find all the regularly self-stable bases
of $\Omega $.

On the other hand, the subfoliations $\hbox{\rm span}\left(
\frac{\partial }{\partial y},\frac{\partial }{\partial z}\right)
$, $\hbox{\rm span}\left( \frac{\partial }{\partial
v},\frac{\partial }{\partial y}\right) $, and $\hbox{\rm
span}\left( \frac{\partial }{\partial v},\frac{\partial }{\partial
z}\right) $ are regularly self-stable $2$-foliations. The first
one is spacelike (its leaves are called {\it wave surfaces}
\cite{Ma}) and the others are lightlike. Regularly self-stable
bases of these subfoliations are given by $\left\{ \frac{\partial
}{\partial y},\frac{\partial }{\partial z}\right\} $, $\left\{
\frac{\partial }{\partial v},\frac{\partial }{\partial y}\right\}
$ , and $\left\{ \frac{\partial }{\partial v},\frac{\partial
}{\partial z}\right\} $ respectively.

Moreover, the timelike $2$-foliation $\hbox{\rm span}\left(
\frac{\partial }{\partial u},\frac{\partial }{\partial v}\right) $
is regularly self-stable too. A regularly self-stable basis is now
given by $\left\{ \frac{\partial }{\partial u},\frac{\partial
}{\partial v}\right\} $.

\section{Discussion and comments}
\label{sec5}

We have introduced some new properties for foliations: stability
and regular stability. Theorem \ref{T1} provides a relationship
between both concepts in terms of the curvature. As particular
cases, self-stability and regular self-stability are two
interesting properties: a self-stable foliation is conserved by
parallel transports along the integral curves of vector fields of
the foliation, and a regularly self-stable foliation has a set of
bases (characterized by Proposition \ref{p100}) whose vector
fields are conserved by parallel transports along the integral
curves of vector fields of the foliation, {\it i.e.} the curvature
of the leaves is ``adapted'' to the curvature of the space-time.
From Corollary \ref{col2} it follows that regular self-stability
is a motion law for flat foliations, in contrast to
self-stability, that it is a motion law for foliations in general.

Finally, we show a direct interpretation of the leaves of a
regularly self-stable lightlike $p$-foliation $\Omega $ with
$p=n-1$, extending some properties of flat wave fronts given in
special relativity (see \cite{Soa,Sob}) to general relativity: let
$\left\{ X_{1},\ldots,X_{p}\right\} $ be a basis of $\Omega $,
where $X_{1},\ldots,X_{p-1}$ are spacelike and $X_{p}$ is
lightlike. Given a world line of a future-pointing timelike vector
field $U$ ({\it i.e.} an observer), the wave fronts of $\Omega $
relative to $U$ are the leaves of the intersection of $\Omega $
and the Landau foliation ${}\mathcal{L}_{U}$ associated to $U$
(see \cite{Lib,Lia,Bo,Oli}). Let $U$ be a future-ponting timelike
vector field such that the wave fronts of $\Omega $ relative to
$U$ are the leaves of the foliation $\hbox{\rm span}\left(
X_{1},\ldots,X_{p-1}\right) $, {\it i.e.}
\begin{equation}
\Omega \cap \mathcal{L}_{U}=\hbox{\rm span}\left(
X_{1},\ldots,X_{p-1}\right) .
\end{equation}
Since $\left\{ X_{1},\ldots,X_{p-1}\right\} $ is a regularly
self-stable basis, the leaves of $\hbox{\rm span}( X_{1},\ldots $
$\ldots ,X_{p-1})$ are totally geodesic and flat. So $U$ observes
the wave fronts of $\Omega $ as spacelike totally geodesic and
flat $\left( n-2\right) $-planes moving in the relative direction
of $X_{p}$ ({\it i.e.} $X_{p}$ projected onto the leaves of
${}\mathcal{L}_{U}$) at the speed of light. But we cannot ensure
that the wave fronts of $\Omega $ relative to any observer are
totally geodesic and flat $\left( n-2\right) $-planes.

\end{document}